\documentclass[aps,twocolumn,prb,floatfix,showpacs]{revtex4}
\usepackage{graphicx}
\usepackage{amsmath}
\usepackage{amssymb}
\newcommand{\MCtwo}{Microtechnology and Nanoscience, MC2, 
Chalmers University of Technology, SE-412 96 G{\"o}teborg, Sweden}

\newcommand{\vdW}{{\mbox{\scriptsize vdW-DF}}}
\newcommand{\Dacapo}{\textsc{Dacapo}}
\newcommand{\vdw}{{\mbox{\scriptsize vdW-DF}}}
\newcommand{\GGA}{{\mbox{\scriptsize GGA}}}
\newcommand{\LDA}{{\mbox{\scriptsize LDA}}}

\newcommand{\nl}{{\mbox{\scriptsize nl}}}

\newcommand{\diff}{ {\rm d}}
\newcommand{\eff}{{\mbox{\scriptsize eff}}}
\newcommand{\ext}{{\mbox{\scriptsize ext}}}
\newcommand{\DFone}{{\mbox{\scriptsize vdW-DF1}}}
\newcommand{\DFtwo}{{\mbox{\scriptsize vdW-DF2}}}
\newcommand{\sfd}{{\mbox{\scriptsize sfd}}}
\newcommand{\Harris}{{\mbox{\scriptsize Harris}}}
\newcommand{\KS}{{\mbox{\scriptsize KS}}}
\newcommand{\nsc}{{\mbox{\scriptsize nsc}}}
\newcommand{\scc}{{\mbox{\scriptsize sc}}}
\newcommand{\self}{{\mbox{\scriptsize self}}}

\begin{document}

\title{A Harris-type van der Waals density functional scheme}

\author{Kristian Berland }\affiliation{\MCtwo}
\author{Elisa Londero}\affiliation{\MCtwo}
\author{Elsebeth Schr{\"o}der}\email{schroder@chalmers.se}\thanks{Corresponding author}\affiliation{\MCtwo}
\author{Per Hyldgaard}\affiliation{\MCtwo}

\date{March 14, 2013}

\begin{abstract} 
Large biomolecular systems, whose function may involve thousands of 
atoms, cannot easily be addressed with parameter-free density functional 
theory (DFT) calculations.  Until recently a central problem was that 
such systems possess an inherent sparseness, that is, they are formed 
from components that are mutually separated by low-electron-density 
regions where dispersive forces contribute significantly to the 
cohesion and behavior.  
The introduction of, 
for example, the van der Waals density functional 
(vdW-DF) method [PRL \textbf{92}, 246401 (2004)] has 
addressed part of this sparse-matter system challenge. 
However, while a vdW-DF study is often as computationally 
efficient as a study performed in the generalized gradient approximation, 
the scope of large-sparse-matter 
DFT is still limited by computer time and memory. It is costly 
to self-consistently determine the electron wavefunctions and 
hence the kinetic-energy repulsion. In this paper we propose 
and evaluate an adaption of the Harris scheme 
[PRB \textbf{31}, 1770 (1985)].
This is done to speed up non-selfconsistent vdW-DF studies
of molecular-system interaction energies.  Also, the 
Harris-type analysis establishes a formal link between 
dispersion-interaction effects on the effective potential 
for electron dynamics and the impact of including 
selfconsistency in vdW-DF calculations 
[PRB \textbf{76}, 125112 (2007)].
\end{abstract}

\maketitle

\section{Introduction} 
Density Functional Theory (DFT) is considered one of the best 
and most reliable condensed-matter tools for non-empirical studies 
of molecular, surface, bulk and compound 
properties.\cite{KieronPerspectives}  Standard implementations, 
using the Local Density Approximation (LDA) or the Generalized 
Gradient Approximation (GGA) for the exchange-correlation 
energy, provide an accurate description of the binding in 
regions characterized by high electron density. An even more 
widespread DFT usage will follow from an ability to address soft- 
and sparse-matter systems, structures that have internal 
voids or low-electron-density regions dominated by
the van der Waals (vdW) forces, also called the London dispersion 
forces.\cite{langrethjpcm2009} While neither 
LDA nor GGA capture the truly nonlocal correlation effects that 
underpin those forces,\cite{Mahan65,Ashcroft,ALL,Tractable,IJQCrev} the 
last decade has seen development of both vdW-extended 
DFT\cite{ALL,ScolesDFTD,HobzaDFTD,Grimme,TS} and of regular nonlocal 
exchange-correlation functionals.\cite{Layered,Dion,JB,Thonhauser,VV09:1,VV09:2,VV10,vdWDF2} 
The first class of methods are often atom centered and require use of a 
damping function or equivalent, while the second class of methods 
fits inside the regular DFT framework. Both types of sparse-matter DFT can 
describe, for example, vdW forces between molecules. 

The van der Waals density functional (vdW-DF) 
method\cite{Tractable,IJQCrev,Dion,Thonhauser,vdWDF2,HRthesis,Dionthesis} 
is a framework for approximating the exchange-correlation energy $E_{xc}[n]$.
The method, summarized below, yields efficient general-purpose sparse-matter 
functionals\cite{Layered,Dion,vdWDF2} that are 
non-empirical. The method employs the Coulomb gauge 
(with Green function $G\equiv |\mathbf{r}-\mathbf{r}'|^{-1}$) 
and a scalar dielectric function $\epsilon$. In its most general 
form,\cite{HRthesis,Dionthesis} the vdW-DF method is a reformulation of 
the adiabatic connection formula\cite{GunnarssonLundqvist,LangrethPerdewACF} (ACF)
and assumes that a plasmon-pole approximation\cite{Tractable,Dion}
for $\epsilon$ can satisfy
\begin{equation}
\int \frac{{\rm d}u}{2\pi} 
\hbox{Tr}\left\{\ln(\nabla \epsilon \nabla G)\right\} \equiv E_{xc}[n] + E_\self,
\label{eq:acfepsilondefine}
\end{equation}
where $u$ denotes the complex frequency and where $E_\self$ 
is the internal Coulomb self-energy
of each electron. 
This equation summarizes an, in principle,  exact description of the 
(longitudinal) electrodynamics in the inhomogeneous electron gas and 
reflects the use of a Dyson equation for handling screening. Like the GGAs, 
the vdW-DF method further uses physics-based constraints\cite{Dion,Thonhauser} 
to approximate the plasmon-pole response and thus defines the functional 
form of a nonlocal correlation term, $E_c^\nl$.  
In the recent explicit functional versions, called 
vdW-DF1\cite{Dion,Thonhauser} and vdW-DF2,\cite{vdWDF2} 
the nonlocal energy $E_c^\nl[n]$ 
is expressed as a double integral over the density, weighted by a kernel. However,
the plasmon basis still allows it to capture a collectivity that 
reflects the broader density variation.\cite{Dion,HRthesis,Dionthesis} 
The vdW-DF method also involves picking a gradient-corrected exchange 
that reflects prespecified criteria, e.g., good all-round 
molecular-system performance\cite{IJQCrev,Dion,rPW86,vdWDF2} 
or improved bulk-system properties.\cite{Cooper09,PBEsol} 
The vdW-DF shares the plasmon-pole emphasis with its LDA and 
GGA relatives, and the functionals ``vdW-DF\#'' have both 
seamless integration in the homogeneous limit and a 
build-in conservation of the exchange-correlation hole.\cite{Dion}  
The non-empirical design suggests 
that the vdW-DFs can achieve a good transferability across 
systems, length scales, charging states, and binding morphologies. 

The vdW-DF method has been and is being tested for many systems.
It delivers a parameter-free atomic-scale 
characterization of the binding in complex sparse-matter 
systems. Selfconsistent (sc) vdW-DF calculations can be 
used to calculate stress within 
periodic unit cells\cite{StressvdWDF} and guide atomic 
optimization (relaxation).  There are performance
tests for bulk,\cite{KGraphite,V2O5a,MolCrys,MolCrys2,V2O5b,Graphane} 
layered\cite{Layered,IJQCrev,ImproveLayered,vdWready} 
absorption,\cite{MOF74CoMg,MOFinterH2,vdWRoleH2uptake} 
molecule and atom 
adsorption,\cite{PAHgraphene,Phenol,SlidingRings,WaterSurfacevdWDF,WetvdWDF,Adenine,vdWNobleGasAdsorb,TiltedAmineUCF,SimplifiedvdW,nalkanes,H2benchmark,vdWswitchRationale,vdWcompareCu,lofwander}
surface self-assembly,\cite{AQassemblyNano,SelfAssemblySi,thioCu} and molecular 
systems.\cite{PolymerCrys,DNAtwist,CNTnature,Cooper09,BenzeneDimerDCL,PAHbindingcompare} 
The development of effective algorithms\cite{Soler,Gulans,Junolo,noloco,nscvdwdf} 
has allowed sc vdW-DF calculations that have today little or 
no additional costs over GGA calculations.\cite{SimplifiedvdW} For very 
large systems, typical of biomolecular-interaction problems, the 
set of real-space vdW-DF evaluation schemes\cite{Gulans,Junolo,noloco,nscvdwdf} 
effectively permits an order-$N$ scaling in the evaluation of 
$E_c^\nl[n]$, as discussed for example in Ref.~\onlinecite{MolCrys2}.

Nevertheless, the feasibility of sparse-matter DFT calculations for 
large molecular matter is still limited by the huge number of atoms that 
are usually involved in interacting molecular systems. The problem is 
confounded by the fact that (especially in biochemically relevant 
problems) we are often forced to address nonperiodic 
systems where size-convergence of the model unit-cell size 
becomes important.\cite{SlidingRings,Adenine} Molecular transport
studies,\cite{Venkataraman} and many  molecular-crystal 
problems,\cite{MolCrys2,TSmolcrysNoa} further exemplify the general need 
to address vdW binding in big systems.
The challenge for typical sparse-matter DFT is 
increasingly becoming that of computing the steric hindrance effects 
that are described by the DFT kinetic-energy repulsion.  When we 
reach thousands of atoms, all first-principle and vdW-extended DFT
calculations are simply limited by memory requirements and 
computational costs of the wavefunction evaluation. 

In this paper we propose and test an adaption of the Harris 
scheme\cite{harris,bell,foulkes,nikulin,gordonkim,footnote} to perform 
non-selfconsistent (nsc) vdW-DF calculations for sparse matter. The 
aim is to explore possibilities for reducing the 
computational cost of the wavefunction-evaluation bottleneck that 
could be impeding an even broader vdW-DF usage in, for example, 
biochemistry.\cite{footnote2}  
The approach consists of using a superposition of
the fragment (electron) densities, and is here termed 
sfd-vdW-DF. It can be seen as 
an alternative to the more costly nsc-vdW-DF or
even sc-vdW-DF evaluations.\cite{IJQCrev,Dion,Thonhauser} 
It is just the regular vdW-DF Harris scheme
if the fragment densities are based on vdW-DF 
calculations. However, the sfd framework also works without a vdW-DF 
implementation of the underlying DFT code, and we reserve the term 
sfd-vdW-DF to cases where the fragment densities are obtained from  
GGA calculations.
The method uses our code for real-space vdW-DF 
evaluation based on the charge densities from the underlying DFT 
code.\cite{CNTnature,KGraphite,MolCrys2,nscvdwdf}

We present formal analysis and a testing of the 
sfd-vdW-DF computational scheme for molecular systems.  
As an interesting aside, the analysis also identifies conditions
for expecting significant vdW-induced changes in the bandstructure 
or electron dynamics for a given binding morphology.
We separately test how well the sfd framework faithfully 
reproduces interaction effects that arise from the semi-local part 
of the vdW-DF functional. We test the performance of the 
sfd-vdW-DF scheme across the S22 benchmark set\cite{S22benchmark} and 
for systems with a varying degree of static polarizations.  
We observe that while a vdW-DF Harris scheme (evaluated 
with vdW-DF fragment densities) is correct to second order in 
binding-induced density changes, $\delta n=n_{\scc}-n_\sfd$, 
the proposed scheme is formally only correct to linear 
order in these density changes.  However, we also find that the 
linear term $\delta n$ is weighted by the changes in the 
effective Kohn-Sham (KS) potential that result from the shift 
from the GGA to the vdW-DF functional.  The  
sfd-vdW-DF scheme may thus be broadly applicable in the absence of 
large static dipoles.

The Paper is organized as follows.  In Section II we summarize
the case for developing an accelerated (but approximate) vdW-DF 
description of biomolecular interaction.  In Section III we present some 
details of the LDA, GGA, and vdW-DF family of constraint-based density 
functionals to facilitate a formal analysis and our proposal for 
the sfd-vdW-DF scheme in Section IV.
Section V provides a brief summary of computational (and implementation)
details. Section VI presents the results of the performance testing
that we have carried out for the proposed sfd-vdW-DF scheme, while Sec.~VII 
contains a discussion.  Finally, Section VIII contains summary and outlook.

\section{A case for fast vdW-DF studies of biomolecular interactions}

In a related work\cite{dnaMapping} we report a vdW-DF mapping 
of the vdW attraction in a DNA dimer (two 
periodic double-helix strands).
That pilot study illustrates that general conclusions 
can be made for the important biomolecular-interaction problem  
from computing merely how $E_c^\nl$ depends on the interaction geometry.

The nonempirical nature and the regular-density-functional basis (no external
parameters) makes vdW-DF well suited to pursue investigations of biomolecular 
interactions.
The full DNA interaction problem touches on two general challenges for refining a computational 
description of life processes. First, it illustrates the workings of molecular recognition 
(the matching by weak forces of the genes  
or just the packing of our genome 
in its environment). Second, it reminds us of the challenge of 
characterizing these effects in a solution that 
contains counter ions.  Since the counter ions, per se, can be expected to play a smaller 
role in the overall interfragment vdW attraction, it is natural to focus a sparse-matter DFT 
study on the charged biomolecular structures themselves. However, for this approach to become 
meaningful, we must be able to also characterize the vdW bonding at various charging states. 
Being a regular (parameter free) nonlocal density functional, vdW-DF has an inherent
advantage for computational studies of biomolecular interactions.

The prospect of vdW-DF studies of biomolecular systems is as promising in
terms of computational cost (and, 
currently, as restrained) as it is for other sparse-matter DFT 
approaches. The evaluation of the nonlocal correlation, $E_c^{\nl}$, 
causes no relevant slow down or bottlenecks thanks to the order-$N$ 
scaling in the real-space evaluation approach for large systems.\cite{MolCrys2} 
In fact, our mapping of DNA attraction\cite{dnaMapping} 
demonstrates that vdW-DF has, in practice, an efficiency similar to that of DFT-D for 
large biomolecular problems. There are for the DNA-dimer attraction 
problem\cite{dnaMapping} no memory bottlenecks and essentially a linear 
scaling at least up to 1000+ cores for the $E_c^{\nl}$ evaluation. 
It is already today possible to bring that functional evaluation to about 
ten minutes wall time.
While this evaluation of $E_c^{\nl}$ does cost more than adding 
the dispersion term in DFT-D,\cite{Grimme} neither of these descriptions 
of the vdW attraction cause limitations: The computing expense is 
irrelevant when compared to the cost 
of converging the density in even one DNA (one periodically repeated coil 
of a DNA double helix) in DFT.

With such a promise, it is frustrating that a full DFT study  of the DNA 
dimer is today impossible without an allocation at a petaflop facility.  
There is a need to accelerate the DFT determination of large-system 
wavefunctions and for sparse matter investigations in general. 
Here we explore an approach  
that focuses on accelerating the evaluation 
of the kinetic-energy repulsion, and which can, in principle, reduce the 
wavefunction-solution stage to just a single 
electronic iteration, as is possible with the regular Harris scheme.\cite{harris,foulkes,HarrisCoding} 

\begin{figure}[t!]
\includegraphics[width=0.45\textwidth]{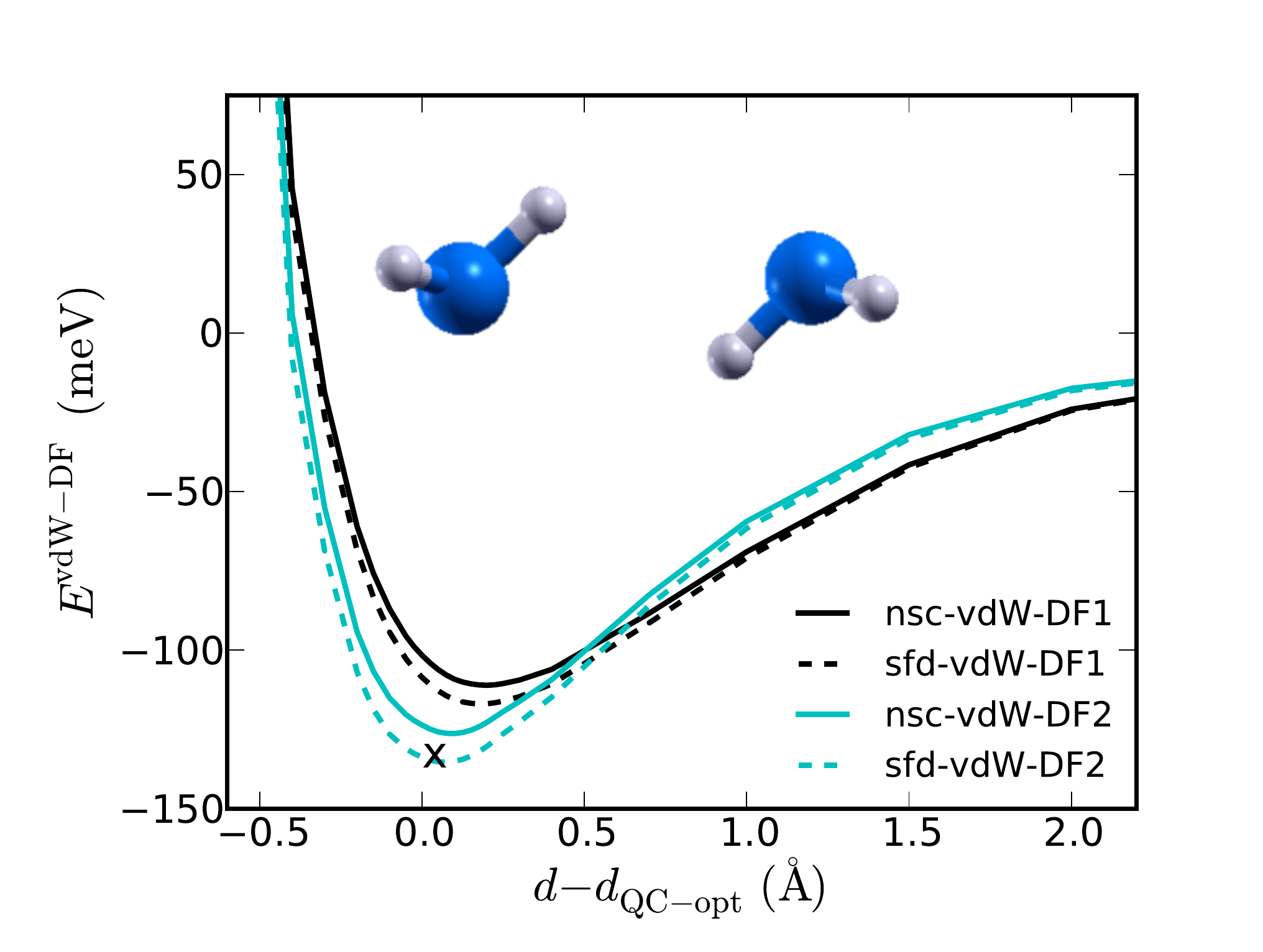}
\caption{
Binding energy curve for the ammonia dimer, 
a member of the S22 benchmark set.\protect\cite{S22benchmark} 
The dark (light) solid curve indicates the results for 
nonselfconsistent vdW-DF1\protect\cite{Dion} (vdW-DF2\protect\cite{vdWDF2,MolCrys2}).
The dashed dark (light) curve indicates the results of   
sfd-vdW-DF1 (sfd-vdW-DF2). 
The cross identifies the binding energy and separation 
as obtained for the original, fully selfconsistent vdW-DF2 
study.\protect\cite{vdWDF2} 
}
\label{fig:S22_dimer}
\end{figure}

\begin{figure}[t!]
\includegraphics[width=0.450\textwidth]{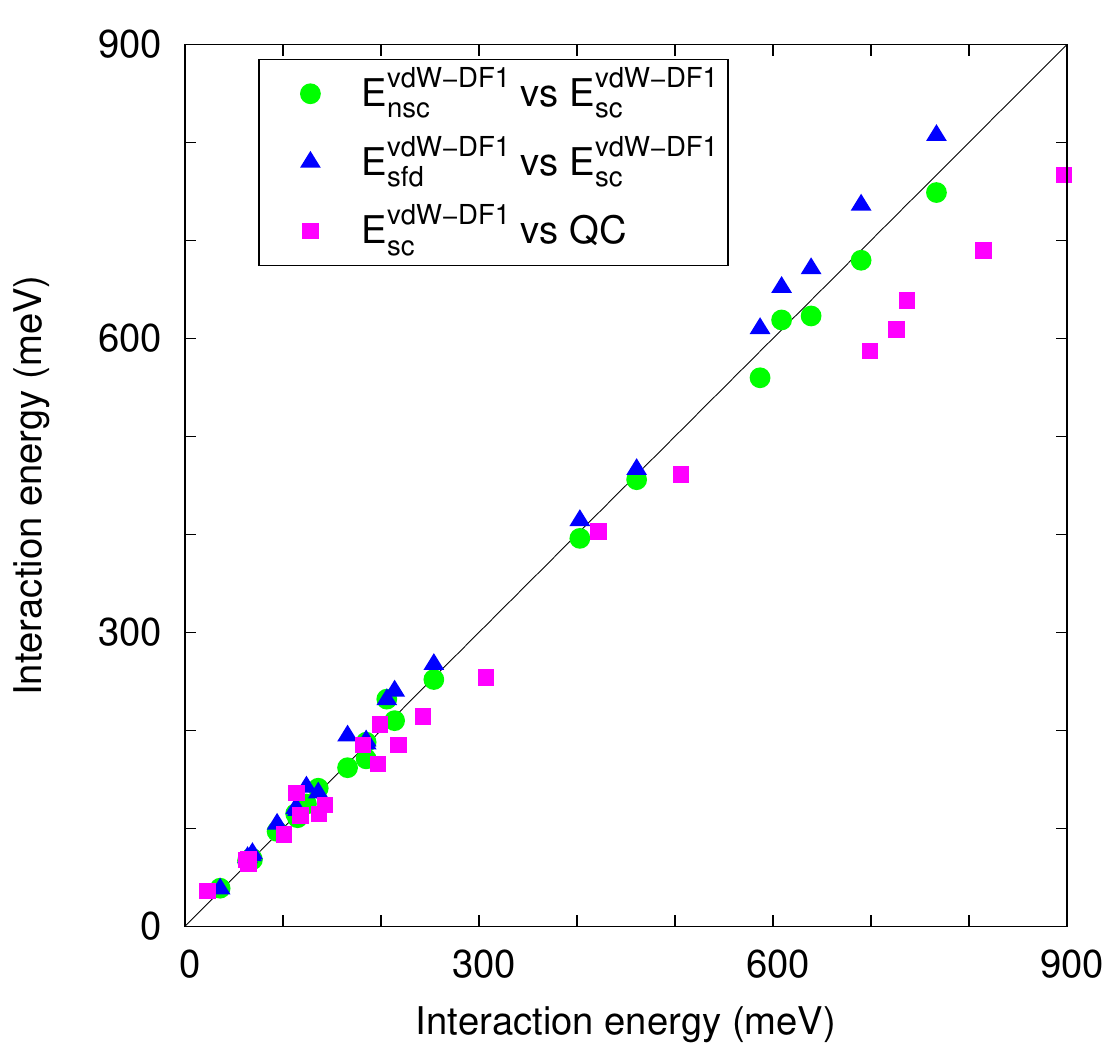}\\
\includegraphics[width=0.450\textwidth]{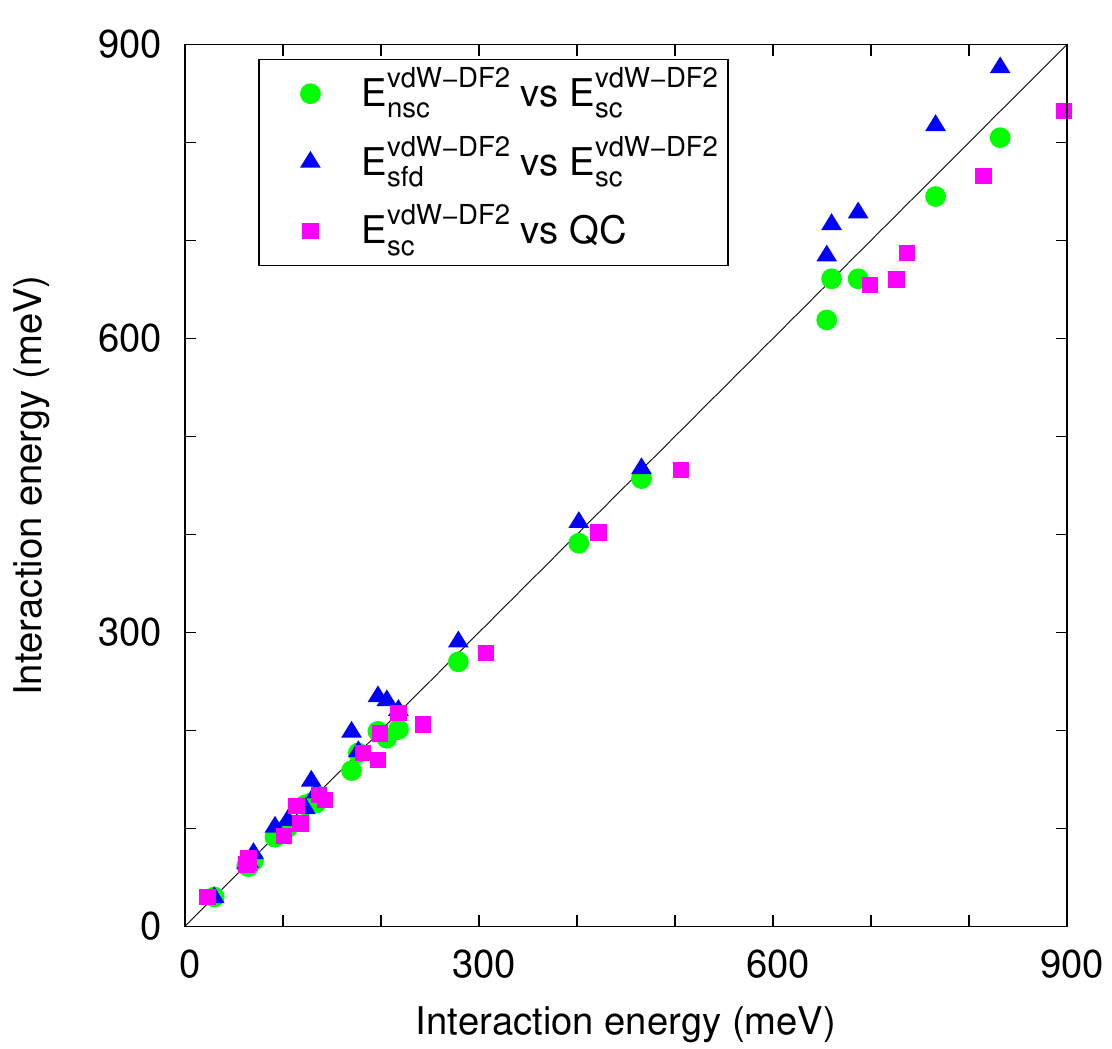}
\caption{
Organic-molecule assessment of the here-described (Harris-type) superposition of fragment densities (sfd) 
evaluation scheme, sfd-vdW-DF.
Results for vdW-DF1 (vdW-DF2) are compared in the top (bottom) panel.
Each point represents a set of binding energies for a pair of molecules in 
the S22 data set. The binding energies are for sfd-vdW-DF and sc-vdW-DF for
vdW-DF1 and vdW-DF2, and as calculated in quantum chemistry (QC) studies.  
The sc and QC results are listed in Ref.~\protect\onlinecite{vdWDF2}, QC results
are originally from Ref.\ \onlinecite{S22QC}.
All sfd and sc calculations are carried out at the 
sc-vdW-DF1 or sc-vdW-DF2 binding distances.\cite{vdWDF2}
We note that the performance of  sfd-vdW-DF  is excellent for S22, 
both as compared to sc-vdW-DF and to QC calculations.
}
\label{fig:milky}
\end{figure}

Figure \ref{fig:S22_dimer} illustrates the feasibility of using the 
sfd-vdW-DF for accelerated vdW-DF studies of biomolecular interactions.
It reports a comparison of binding in an ammonia dimer and shows that 
the sfd-vdW-DF2 description is in excellent agreement with the binding
predicted by sc-vdW-DF2 (indicated by a cross). 

Figure \ref{fig:milky} summarizes our overall assessment, further
detailed in Sec.~VI. It testifies to a high degree of robustness across 
the S22 benchmark set\cite{S22benchmark} and for other molecular-interaction 
problems. The important observation for developing a biomolecular
computational strategy in DFT is that the sfd-vdW-DF scheme can be as 
reliable as the often-used nsc-vdW-DF 
evaluation\cite{Dion,CNTnature,KGraphite,MolCrys2} even if it bypasses 
the need for a sc determination of the wavefunctions. 
The sfd framework 
could thus be one approach to speed 
up vdW-DF studies of large biomolecular interaction problems at a 
limited cost in accuracy.

The (regular) Harris scheme works within 
a fixed functional choice and takes the following steps: 
First, self-consistent calculations of the densities 
$n_{i=1,2,\ldots}$ for each individual building block `$i$'; 
second, construction of a density $n_\sfd=\sum_i n_i$ as a 
superposition of the building blocks,  and of the effective 
single-particle potential $V_{\eff,\sfd}(\mathbf{r})=V_\eff[n_\sfd](\mathbf{r})$ 
that corresponds to $n_\sfd$; and third a nsc, that 
is, no-density update, calculation of the eigenvalues corresponding to 
the potential $V_{\eff,\sfd}(\mathbf{r})$. These eigenvalues help provide an estimate of
the kinetic energy term in the Harris scheme.\cite{harris} 
The Harris scheme is 
traditionally pursued in a LDA or GGA framework starting, e.g., from sc 
GGA input densities. It is today often  used when pursuing bandstructure calculations 
in DFT (and therefore provides very accurate wavefunctions for a given 
$V_{\eff,\sfd}(\mathbf{r})$).
However, it is also possible to use it for its original purpose, namely for providing
efficient but approximate DFT studies of interactions.

Pursuing a regular vdW-DF Harris scheme 
represents one alternative for accelerating vdW-DF studies.  We point out that the 
original study by Harris\cite{harris} (working with LDA) shows that the scheme works 
reasonably well also for describing the formation of covalent bonds between 
atoms in some molecules. It should be even better suited 
to describe the weaker vdW bonding.
The overall criteria for the applicability of a Harris-type 
scheme is that the fully sc density solution $n_{\scc}$ should 
not differ significantly from the input superposition density $n_\sfd$.  
It can work also when we consider systems, like biomolecules in solution, 
where the charging from the surrounding counter ions must be considered 
when providing the input densities (and hence $n_\sfd$ and
$V_{\eff,\sfd}(\mathbf{r})$).  A vdW-DF Harris scheme can be expected to work well for
a study of supramolecular systems as long as the \textit{binding-induced\/} 
charge relocations remain small. 

A regular Harris vdW-DF scheme is, however, not the focus 
here. This is because we do not yet have the ability to both perform vdW-DF 
calculations and allow an externally-defined superposition of input density 
in the same code.  
We have chosen to build 
on the \textsc{Dacapo}\cite{dacapo} code (for input densities and for the nsc 
kinetic-energy evaluation\cite{HarrisCoding}) and on the vdW-DF 
postprocessing\cite{CNTnature,MolCrys2} that we have previously used extensively for 
vdW-DF studies.
A benefit of introducing sfd-vdW-DF as a supplement to a regular vdW-DF
Harris scheme is 
that we thus provide a computational framework that 
can take input densities from an arbitrary code (through an adaption 
of any of the real-space strategies for evaluation 
vdW-DF\cite{Gulans,Junolo,noloco,MolCrys2,nscvdwdf}).

\section{Density functional descriptions of dense and sparse materials}

We begin with a description of differences between the regular form of the nonlocal functional 
vdW-DF and the traditional local (LDA) and semilocal (GGA) DFT descriptions. This  
facilitates our subsequent discussion of the approximation that can allow an accelerated evaluation 
in (nsc) vdW-DF studies.

\subsection{Nonempirical approximations to the universal energy density functional,
LDA, GGA, vdW-DF}

Both GGA and vdW-DF are refinements of the earlier approach LDA that 
describes the universal exchange correlation functional
\begin{eqnarray}
	E_{xc}^{\LDA}[n] &=& E_x^{\LDA}[n]+ E_c ^{\LDA}[n] \\
  E^{\LDA}_{x,c}[n] & = & \int {\rm d}^3\mathbf{r}\ n(\mathbf{r}) 
	\varepsilon_{x,c}^{\LDA}(n(\mathbf{r}))
\end{eqnarray}
in terms of exchange and correlation energy densities 
$\varepsilon_{x,c}^{\LDA}(n(\mathbf{r}))$ that are just functions 
of the local density $n(\mathbf{r})$. The behavior of the LDA was 
initially established by considering the self-energy shifts that 
result with a single-particle coupling to the collective plasmon 
excitations.\cite{GunnarssonLundqvist,LangrethPerdewACF} The LDA description 
was later refined by considering Quantum Monte Carlo studies of the 
homogeneous electron gas and its response.\cite{CAqmc,VWN,PW92} 

The GGA adds a functional dependence on the local gradient through 
dimensionless measures of the gradient, 
$s(\mathbf{r})=|\nabla n|/(2 k_F n)$ and 
$t(\mathbf{r})\propto|\nabla n|/(n k_{s})$, 
where $k_F=(3 \pi^2 n)^{1/3}$ is the Fermi wave number, 
$k_{s}=(4 k_F/\pi a_0)^{1/2}$ denotes the Thomas-Fermi screening wave 
number, and $a_0=\hbar^2/me^2$. 
The set of GGAs expresses the exchange-correlation energy
\begin{equation}
E_{xc}^{\GGA} [n]\equiv 
\!\!\int\!\! \diff^3 \mathbf{r}\, n(\mathbf{r}) 
[\varepsilon_x^{g}(n(\mathbf{r}),s(\mathbf{r}))\!+
\!\varepsilon_{c}^{g}(n(\mathbf{r}),t(\mathbf{r}))];
\label{eq:ggaform}
\end{equation}
we keep a subscript `$g$' on the GGA energy densities to stress
that one must pick a particular design choice, although 
history has pulled towards a few major choices.\cite{KieronPerspectives} 

The inclusion of the dimensionless gradient permits a richer variation
of functional forms. 
The set of constraint-based GGAs is among the most 
successful\cite{KieronPerspectives} 
and these extend the plasmon picture of the LDA via a wavevector 
analysis,\cite{LangrethPerdewACF,LangrethMehl,Perdew} while also emphasizing
conservation of the exchange-correlation hole. The development led to
robust and very versatile GGA forms like the PBE.\cite{PBE} In the 
constraint-based GGAs, the form of the exchange energy density
\begin{equation}
\varepsilon_x^{g}(n(\mathbf{r}),s(\mathbf{r})) = 
\varepsilon_x^\LDA(n(\mathbf{r})) F_x^g(s(\mathbf{r})),
\end{equation}  
is given by an enhancement factor $F^g_{x}(s(\mathbf{r}))$ that has been 
chosen to satisfy a number of scaling laws, formal constraints and 
guidelines.\cite{LangrethMehl,Perdew,ElliotBurke,PBE}
Formal analysis also guides the choice of the gradient-corrected 
correlation energy density $\varepsilon_{c}^{g}(n(\mathbf{r}),
t(\mathbf{r}))$. It is important to note that the richer variation 
that (\ref{eq:ggaform}) supports is thus tempered by adherence to 
fundamental physics criteria and that the identification and use 
of these criteria has helped propel the GGA (and DFT) to a 
tremendous success.\cite{KieronPerspectives}

The broad class of materials and systems that are characterized 
by sparseness does, however, require further refinements 
beyond GGA. For example, organics, biomatter, and supramolecular 
systems are sparse in the sense that they have internal electron 
voids or other regions with a low electron distribution. Here the
binding and function are dominated by the vdW forces
that reflect an electrodynamic coupling and act across internal 
voids. 

The vdW-DF method goes beyond LDA and GGA by introducing a truly 
nonlocal correlation contribution $E_c^{\nl}[n]$ that makes the 
electrodynamical coupling explicit\cite{Tractable,Layered,Dion,vdWDF2} 
via Eq.~(\ref{eq:acfepsilondefine}).  The vdW-DF 
exchange-correlation energy is thus written
\begin{equation}
	E_{xc}^{\vdw}[n] = E_{xc}^{v,0}[n]+E_c^\nl[n]
	\label{eq:fullExcvdWDF}
\end{equation}
where 
\begin{equation}
E_{xc}^{v,0}[n] = E_{c}^{\LDA}[n] + E_x^{v}[n] 
\label{eq:semilocfunc}
\end{equation}
denotes the semilocal part of the functional. We use a superscript `$v$'
to identify the vdW-DF versions\cite{Layered,Dion,vdWDF2} and stress that
these in general have different exchange components $E_x^{v}[n]$. We note 
that the different vdW-DF versions also have different forms of 
the nonlocal correlation term but we have chosen not to make that 
explicit in our discussion.

In the recent vdW-DF versions the nonlocal correlation term is
expressed by a second-order expansion in the plasmon-pole 
response\cite{Dion,vdWDF2}
\begin{equation}
E_c^\nl [n] = \frac{1}{2} \int {\rm d}^3 \mathbf{r} \int {\rm d}^3\mathbf{r}' \,
n(\mathbf{r}) \phi[n](\mathbf{r},\mathbf{r}') n(\mathbf{r}')
\,.
\label{eq:EcnlvdWDF}
\end{equation}
This nonlocal correlation term still captures the broader density 
variation through a collectivity that the plasmon reflects and
(\ref{eq:EcnlvdWDF}) is designed so that the vdW-DF method avoids 
double counting with the terms captured in the
local correlation. The vdW-DF\# have a seamless integration
\begin{equation}
E_c^{v}[n] = E_c^\LDA[n] +E_c^\nl [n], 
\end{equation}
and thus bypass the need for using a damping function. As stressed in 
the introduction, the vdW-DF\# are derived as an approximation to the ACF.
The recent more explicit functionals also use the ACF to link the plasmon pole 
to an inner functional\cite{HRthesis,Dionthesis,Dion,Thonhauser,vdWDF2} that 
is also of the form (\ref{eq:semilocfunc}).  All functionals in the vdW-DF method 
build the nonlocal functional from the inside out, i.e., describe the 
electrodynamics of the dispersion forces by linking to response of
the time-tested LDA/GGA plasmon description.

In this paper we use the vdW-DF method and work with both the vdW-DF1 
version (in which the revPBE \cite{revPBE} GGA is used for 
$E_x^{\vdw}[n]$) and the vdW-DF2 version (in which the re-fitted 
PW86 \cite{PW86,rPW86} GGA is used for $E_x^{\vdw}[n]$). 
We note that in addition to the canonical Rutgers-Chalmers vdW-DF 
versions\cite{Layered,Dion,Thonhauser,Cooper09,vdWDF2} there are
also variants.\cite{optPBE,optBEEFs} These variants fit  
the outer exchange functional $E_{x}^{\vdw}[n]$ to a form that is
fitted to specific data sets, for example the S22.

\subsection{The exchange-correlation potentials}

For a discussion of the nature of the Harris and nsc-vdW-DF schemes (below)  
it is important to also characterize differences 
in the corresponding exchange-correlation potentials
\begin{eqnarray}
\mu_{xc}^{g}[n](\mathbf{r}) & \equiv 
& \frac{\delta E_{xc}^{g}[n]}{\delta n(\mathbf{r})},\\
\mu_{xc}^{v}[n](\mathbf{r}) & \equiv 
& \frac{\delta E_{xc}^\vdw[n]}{\delta n(\mathbf{r})}.
\end{eqnarray}
We also use these potentials in a discussion of the error in the sfd-vdW-DF.
Again we have used superscripts $g$ and $v$ to stress that for calculations
we must pick specific versions of the GGA or of the vdW-DF.

Ref.~\onlinecite{Thonhauser} provides a derivation of the 
effective-potential contribution $\Delta \mu_c^\nl[n](\mathbf{r})$ that 
arises from taking functional derivatives of the nonlocal correlation 
term $E_c^\nl$.  From (\ref{eq:fullExcvdWDF}) it follows that 
the vdW-DF exchange-correlation potential will also differ from 
a GGA exchange-correlation potential 
\begin{eqnarray}
\lefteqn{\mu_{xc}^{v}[n](\mathbf{r})-\mu_{xc}^{g}[n](\mathbf{r}) =} 
\nonumber \\[0.7em]
&& \Delta \mu_{xc}^{v,0}[n](\mathbf{r})+ \Delta \mu_c^\nl[n](\mathbf{r})
\end{eqnarray}
by a semilocal potential term $\Delta \mu_{xc}^{v,0}[n]$. This 
semilocal potential change arises in part because vdW-DF 
subtracts off the gradient corrections to correlation. Also 
when discussing the difference from a given GGA `$g$' 
of exchange-energy form $E_x^{g}[n]$, the semilocal 
potential change must reflect the differences 
$E_x^{v}[n]-E_x^{g}[n]$.

\subsection{Selfconsistent DFT calculations of the total energy}

The KS 
 energy can be written\cite{harris}
\begin{equation}
E_{\KS} [n]= T_0 +
\int d^3\mathbf{r} \, n(\mathbf{r}) 
\left[\frac{1}{2}\phi_n(\mathbf{r}) +V_\ext(\mathbf{r})\right]
+ E_{xc}[n] + E_N.
\label{eq:EKS}
\end{equation}
Here $V_\ext(\mathbf{r})$ is the external potential and 
$\phi_n(\mathbf{r})$ the electrostatic potential at the given 
density $n(\mathbf{r})$,
\begin{equation}
\phi_{n}(\mathbf{r}) = 
\int d^{3}\mathbf{r}' \frac{n(\mathbf{r}')}{|\mathbf{r}-\mathbf{r}'|}.
\end{equation}
The first term $T_0$ and the last term $E_N$ of (\ref{eq:EKS}) express 
the kinetic energy of an effective single-particle wavefunction 
problem and the internuclear repulsion term, respectively.

Fully selfconsistent DFT calculations in the KS scheme 
proceed by finding the single-particle wavefunctional 
solution of an effective eigenvalue problem
\begin{equation}
\left\{-\frac{1}{2} \nabla^2 + V_\eff [n] (\mathbf{r})
- \epsilon_\lambda\right\} \psi_\lambda(\mathbf{r}) = 0
\label{eq:SPKSeigenvalues} 
\end{equation}
defined by the density-dependent (and density-functional specific) effective potential
\begin{equation}
V_\eff[n](\mathbf{r}) \equiv V_\ext(\mathbf{r}) + \phi_n(\mathbf{r}) + \mu_{xc}[n](\mathbf{r}).
\label{eq:effectivepotential} 
\end{equation}
We use atomic units in all formal discussions of DFT calculations and 
the set of approximations.
Selfconsistency is enforced by requiring that the resulting single-particle 
description of the electron density 
\begin{equation}
\tilde{n}(\mathbf{r}) = \sum_\lambda^{\rm occ} |\psi_\lambda(\mathbf{r})|^2
\label{eq:SPdensitydescr}
\end{equation}
actually coincides with the density that specified the effective 
single-particle potential (\ref{eq:SPKSeigenvalues}).

As an example of the total-internal energy DFT calculation that is 
thus made possible, we consider a two-fragment system with components 
separated by a distance $d$.  We here follow the presentation in 
Ref.~\onlinecite{harris} so as to simplify
our subsequent discussion (Section IV). Summing up the set of occupied, 
single-particle energies $\epsilon_{\lambda}$, leads to an incorrect counting 
of the total electron-electron interaction energy.  However, this complication
is easily adjusted for, giving\cite{footnote3}
\begin{eqnarray}
E_{\KS}(d) & = & \sum_{\lambda}^{\rm occ} \epsilon_\lambda -  
\int d^3 \mathbf{r} \,n(\mathbf{r})\,\left\{ \frac{1}{2}\phi_{n_{\scc}}(\mathbf{r}) + 
\mu_{xc}[n_{\scc}](\mathbf{r})\right\}
\nonumber\\
& & {}+ E_{xc}[n_{\scc}] +E_N(d).
\end{eqnarray}
All terms depend on the mutual fragment separation $d$ (although
we do not make that an explicit statement for all terms).  We use 
$E_{\KS}^{\vdw}(d)$ or $E_{\KS}^{g}(d)$ to specify whether
the sc DFT total energy result was pursued in 
a vdW-DF or a GGA choice, respectively.
Below we focus the discussion on such fragment problems with mutual 
separation $d$.

\subsection{Non-selfconsistent vdW-DF calculations}

We present an overview of the standard nsc-vdW-DF evaluation\cite{Dion,langrethjpcm2009} 
which has 
a total-energy variation $E^{\vdw}_{\nsc}(d)$. This energy variation is 
often\cite{Thonhauser,langrethjpcm2009} found to be a close 
approximation to the energy variation $E_{\scc}^{\vdw}(d)
\equiv E_{\KS}^{\vdw}(d)$ found by fully 
sc vdW-DF calculations, Sec.~III.C.

The nsc-vdW-DF evaluations proceed for a given GGA 
`$g$' by first completing self-consistent GGA calculations of both the 
electron density variation $n^{g}_{\scc}(d)$ and total GGA internal 
energy $E_{g}(d)$. The GGA choice $g$ is in practice often PBE and perhaps more 
seldom the revPBE version (that vdW-DF1 uses for its exchange component 
but that is of no concern in 
this formal discussion). We denote by $E_{xc}^{g}(d)$ the corresponding
exchange and correlation energy that are evaluated for $n^{g}_{\scc}(d)$.
The nsc-vdW-DF calculations proceed by simply adjusting 
for the nonlocal correlation and for the differences in semi-local 
correlation terms
\begin{equation}
E^{\vdw}_{\nsc}(d)=
E^{g}(d) + E_c^{\nl}[n_{\scc}^{g}(d)]+\Delta E_{xc}^{v,0} [n_{\scc}^{g}(d)].
\label{eq:nscvdWDF}
\end{equation}
The semilocal functional component
\begin{equation}
\Delta E_{xc}^{v,0} [n_{\scc}^{g}]=
E_c^\LDA [n_{\scc}^{g}] + E_x^{v} [n_{\scc}^{g}] - E_{xc}^{g}(d) 
\end{equation}
not only extracts the gradient corrections to correlations but also 
implements a possible adjustment in the gradient-corrected exchange 
description (as necessary).

The nsc-vdW-DF calculations were for some time the 
only manner for completing a vdW-DF study: It permitted us to 
include van der Waals interactions  in a computational 
efficient parameter-free single-density functional DFT.\cite{Layered,Dion,Thonhauser}
The approach can 
be motivated, in part by a Harris-type description (as 
substantiated further below) but the quantitative extent of the 
approximation could only be tested when efficient implementations of 
the sc-vdW-DF method were made available.\cite{Thonhauser,Soler} 
The subsequent testing showed that nsc-vdW-DF often captures most of  
the binding of sc-vdW-DF.\cite{Thonhauser}

\section{Harris-type evaluation schemes}

This paper formally proposes a computational strategy that combines 
nsc-vdW-DF calculations (above) with a further approximation 
inspired by the Harris scheme\cite{harris,foulkes} and other earlier
suggestions of using frozen fragment densities.\cite{nikulin,gordonkim}
The approximation can limit the computational costs for molecular systems 
because it reduces the quality required for the input 
density in the nsc-vdW-DF evaluation. It comes with an 
accuracy cost, which as expected is found largest for systems with a 
static polarization, but it can provide a speed up.

\subsection{Variational nature of the Harris scheme}

The regular Harris scheme rests ultimately on the variational character of 
the KS formulation of the total energy for fully selfconsistent DFT evaluations. 
The KS energy form acquires a minimum at the correct ground-state density $n_{\scc}$,
\begin{equation}
E_{\KS}[n_{\scc}+\delta n] = E_{\KS}[n_{\scc}] + C(\delta n)^2; \quad C > 0.
\end{equation}

The Harris scheme rewrites the KS formulation of the total energy so as 
to avoid the need for updating the electron density in an estimate of the interacting 
energy. As mentioned in the introduction, the Harris scheme does not provide nor 
does it work 
with the sc density $n_{\scc}$, but rather utilizes a superposition 
density 
\begin{equation}
n_{\sfd} \equiv \sum_i n_{{\scc},i}
\end{equation}
defined from sc determinations of the electron
densities $n_{{\scc},i}$ for each of the fragments of the weakly interacting system.

In a GGA study, for example, we can formally express the Harris 
interaction estimate\cite{harris}
\begin{eqnarray}
E^{g}_{\Harris}(d)&\equiv&\sum_{\lambda}^{\rm occ} \bar{\epsilon}_{\lambda}^g\nonumber \\
&&{}-\int d^3\mathbf{r} \; 
n_{\sfd}^g(\mathbf{r})\left\{\frac{1}{2}\phi_{n_{\sfd}^g}(\mathbf{r})+\mu_{xc}^{g}[n_{\sfd}^g](\mathbf{r})\right\}\nonumber \\
&&{}+E_{xc}^g[n_{\sfd}^g]+E_N(d).
\label{eq:HarrisGGA}
\end{eqnarray}
Again, $d$ is the distance between the fragments and $\phi_{n_\sfd^g}$ 
is the electrostatic potential defined by $n_\sfd^g$.  In the Harris 
estimate (\ref{eq:HarrisGGA}), the eigenvalues $\bar{\epsilon}_\lambda^g$
are the single-particle energies calculated within the Harris ``one-shot" 
(no density update) GGA calculation for the frozen input superposition 
density $n_{\sfd}^g$ using the effective (Harris-GGA) potential 
\begin{equation}
V_{\eff,\sfd}^g(\mathbf{r})= V_\eff^g[n_{\sfd}^g](\mathbf{r})=
V_\ext(\mathbf{r})+ \phi_{n_{\sfd}^g}(\mathbf{r})+\mu_{xc}^g[n_{\sfd}^g](\mathbf{r}).
\label{Veffharrisg}
\end{equation}
There does, of course, exist a corresponding expression for a Harris approximation
to vdW-DF calculations, in which case the input density would be $n_{\sfd}^{\vdw}=
\sum_i n_{{\scc},i}^\vdw$.  

The central step in the Harris scheme is the assumption that the density change 
$n_{\scc}-n_{\sfd}$ produces only a small change in the effective potential,
\begin{eqnarray}
\Delta V_\eff(\mathbf{r}) & = & \phi_{n_{\scc}}(\mathbf{r}) 
- \phi_{n_{\sfd}(\mathbf{r})}  \nonumber \\
& & {}+ \mu_{xc}^g[n_{\scc}](\mathbf{r}) -\mu_{xc}^g[n_{\sfd}](\mathbf{r}),
\end{eqnarray}
so that one can expand the difference in KS and Harris estimates for the single-particle energy sum
\begin{equation}
\sum^{\rm occ}_{\lambda} 
\epsilon_{\lambda} - 
\sum^{\rm occ}_{\lambda'} 
\bar{\epsilon}_{\lambda'} 
= \int d^3\mathbf{r} \, n_{\scc}(\mathbf{r})
\Delta V_{\eff}(\mathbf{r}) + {\cal O}(n_{\scc}-n_{\sfd})^2.
\end{equation}
The linear term cancels out corresponding linear terms in the expansion of $E_{xc}[n]$ and in the
calculation of the electrostatic potential.\cite{harris}

The resulting single-shot (no density update) DFT estimate is also variational 
\begin{equation}
E_{\Harris}[n_{\sfd}] = E_{\Harris}[n_{\scc}] +{\cal O}(n_{\scc}-n_{\sfd})^2, 
\label{eq:HarrisExpand}
\end{equation}
but it is not, in general, an extremum.\cite{harris,foulkes,zaremba,Farid} 
This follows because there is no consistency 
between the Harris scheme input density $n_{\sfd}$, and 
the single-particle electron density (\ref{eq:SPdensitydescr}) that 
results with the Harris-scheme effective potential 
$V_{\eff}[n_{\sfd}](\mathbf{r})$. 

\subsection{Nature of and error in non-selfconsistent vdW-DF calculations}

The non-selfconsistent vdW-DF total energy $E_{\nsc}^{\vdw}(d)$ is an 
approximation to the fully selfconsistent vdW-DF result 
$E_{\scc}^{\vdw}(d)\equiv E_{\KS}^{\vdw}(d)$.
However, unlike a regular vdW-DF Harris interaction estimate 
$E_{\Harris}^{\vdw}(d)$ it is built from the sc GGA result for the entire 
system $n_{\scc}^g$ and not the superposition of sc-vdW-DF fragment 
densities, $n_{\sfd}^{\vdw} = \sum_i n_{{\scc},i}^{\vdw}\neq n_{\scc}^g$. 
The nsc-vdW-DF approximation $E_{\nsc}^{\vdw}$ can, however, still 
formally be seen as a further extension of the ideas that underpin 
the Harris estimate Eq.~(\ref{eq:HarrisGGA}). 

To establish a formal relation between sc and nsc vdW-DF calculations, 
we consider the differences in sc results that arise as we replace a 
GGA choice $g$ with a vdW-DF choice for the exchange-correlation 
functional. We introduce
\begin{eqnarray}
\delta n_{\scc} & \equiv & n_{\scc}^\vdw-n_{\scc}^g,\\
\Delta V_{\scc} & \equiv & \phi_{n_{\scc}^{\vdw}}-\phi_{n_{\scc}^g}
\nonumber\\
& & {}+ \mu_{xc}^\vdw[n_{\scc}^\vdw]-\mu_{xc}^g[n_{\scc}^g] ,
\end{eqnarray}
to identify the changes resulting in the density and in the effective 
potential, respectively.
As in the original Harris analysis,\cite{harris} 
we can consider both $\delta n_{\scc}$ and $\Delta V_{\scc}$ 
small, and thus giving rise only to linear changes 
\begin{eqnarray}
\sum_{\lambda}^{\rm occ} \epsilon_\lambda^\vdw
-\sum_{\lambda}^{\rm occ} \epsilon_\lambda^g & \approx & 
\int d^3\mathbf{r}\, n_{\scc}^{\vdw}(\mathbf{r})\Delta V_{\scc}(\mathbf{r}),\\
E_{xc}[n_{\scc}^{\vdw}]
-E_{xc}[n_{\scc}^{g}] & \approx & 
\int d^3\mathbf{r}\, \delta n_{\scc}(\mathbf{r})
\mu_{xc}^{\vdw}[n_{\scc}^g](\mathbf{r}).\nonumber\\
\end{eqnarray}

Simply extending the analysis behind the Harris estimate therefore
yields the formal relation
\begin{eqnarray}
E^{\vdw}_{\scc}(d) &\approx & E^{g}_{\KS}(d) + E_c^\nl[n_{\scc}^g]+
\Delta E_{xc}^{\vdw,0}[n_{\scc}^g]\nonumber\\
&& {}+\Delta E_{\mu, {\scc}}^{\vdw\leftarrow g} + 
{\cal O}(\delta n_{\scc})^2\\
& = & E^{\vdw}_{\nsc}(d) + \Delta E_{\mu, {\scc}}^{\vdw\leftarrow g} + 
{\cal O}(\delta n_{\scc})^2\nonumber \\
\label{eq:scvdWDFexpress}
\end{eqnarray}
with leading-order correction term
\begin{eqnarray}
\lefteqn{\Delta E_{\mu, {\scc}}^{\vdw\leftarrow g}=}\nonumber \\ 
&  & \int\!\! d^3\mathbf{r}\,\delta n_{\scc}(\mathbf{r})
\left\{\mu_{xc}^\vdw[n_{\scc}^\vdw](\mathbf{r})
-\mu_{xc}^{g}[n_{\scc}^g](\mathbf{r})\right\}.
\label{eq:densitynscvdWDFexpress}
\end{eqnarray}

\subsection{A sfd-vdW-DF scheme for accelerated calculations of 
molecular interactions}

We propose to pursue a Harris-type
vdW-DF scheme that is based on the superposition $n_\sfd^g=n_1^g+n_2^g+\ldots$ 
of GGA fragment densities but which approximates the vdW-DF total energy by
\begin{equation}
E_{\sfd}^{\vdw}(d) \equiv E_{\Harris}^g(d) + E_c^{\nl}[n_{\sfd}^g]
+\Delta E_{xc}^{\vdw,0}[n_{\sfd}^g].
\label{HtypevdWDFDefined}
\end{equation}
Here, again, $E_{\Harris}^g(d)$ denotes the regular Harris estimate as described 
in the given GGA choice $g$.

A formal analysis motivating the proposed vdW-DF approximation 
(\ref{HtypevdWDFDefined}) is essentially already stated in Section III.B. 
We now consider slightly different density 
and effective-potential differences
\begin{eqnarray}
\delta n_{\sfd} & \equiv & n_{\sfd}^{\vdw}-n_{\sfd}^g,\\
\bar{\Delta V}_{\sfd} & \equiv & \phi_{n_{\sfd}^{\vdw}}-\phi_{n_{\sfd}^g}
\nonumber\\
& & {}+ \mu_{xc}^{\vdw}[n_{\sfd}^{\vdw}]-\mu_{xc}^g[n_{\sfd}^g],
\end{eqnarray}
so that the $g\to \vdw$ changes instead reflect the effects on the Harris-scheme 
single-particle eigenenergies $\bar{\epsilon}_{\lambda}$.

The sfd-vdW-DF estimate can thus be expressed as an approximation
to a regular vdW-DF Harris scheme
\begin{equation}
E^{\vdw}_{\Harris}(d) 
= E^{\vdw}_{\sfd}(d) + \Delta E_{\mu, {\sfd}}^{\vdw \leftarrow g} + 
{\cal O}(\delta n_{\scc})^2
\label{eq:HnscvdWDFexpress}
\end{equation}
where we now have a slightly different leading-order correction term
\begin{eqnarray}
\lefteqn{\Delta E_{\mu, {\sfd}}^{{\vdw}\leftarrow g}=}\nonumber\\ 
& & \int\!\! d^3\mathbf{r}\,\delta n_{\sfd}(\mathbf{r})
\left\{ 
\mu_{xc}^{\vdw}[n_{\sfd}^{\vdw}](\mathbf{r})
-\mu_{xc}^{g}[n_{\sfd}^g](\mathbf{r})
\right\}.
\label{eq:densityHnscvdWDFexpress}
\end{eqnarray}

\section{Computational details}

This paper compares the interaction energy curves obtained with nsc 
Harris-type calculations with those of DFT calculations for selected 
non-covalently bound molecular dimers. Both sc and nsc calculations with 
the PBE version\cite{PBE} of GGA are performed using the \textsc{Dacapo} 
software.\cite{dacapo}
This planewave DFT code was chosen because it is straightforward in \textsc{Dacapo} to set 
the electronic densities equal to the sum of molecular (frozen input) 
densities $n_{\sfd} = n_1 + n_2$ through an external manipulation
in \textsc{ase}\cite{ase} and thus to prepare the sfd calculations.

The non-local correlation energy is evaluated in a post-processing 
procedure both for regular DFT and Harris-type vdW-DF calculations. 
For these calculations, we use an efficient in-house real-space code, 
further described in Ref.~\onlinecite{MolCrys2}. 
A radius cutoff of 
$6$ {\AA} is used for dense (full) sampling of the grid
and a cutoff of $26$ {\AA} is used for sparse (double-spaced) sampling of the grid. 

In the PBE calculations, relying on Vanderbilt ultrasoft pseudopotentials,  
we use plane-wave and density-sampling cutoffs of 500 eV.
This cutoff choice has been used in many similar 
calculations\cite{MolCrys,MolCrys2,Adenine} and gives a relatively dense 
sampling of the density grid used to evaluate the non-local correlation.
As long as the reference calculations have the same grid-sampling density, 
here secured by using the same size of the unit cell, 
the non-local correlation energy is typically converged to within about 1~meV. 

\section{Results: assessing the sfd-vdW-DF evaluation}

\begin{figure}[t!]
\includegraphics[width=0.43\textwidth]{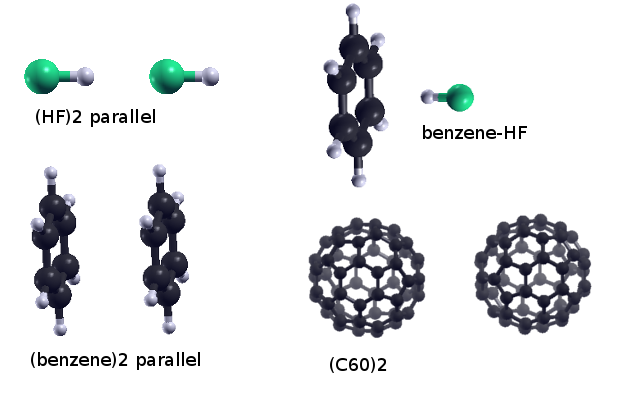}
\includegraphics[width=0.43\textwidth]{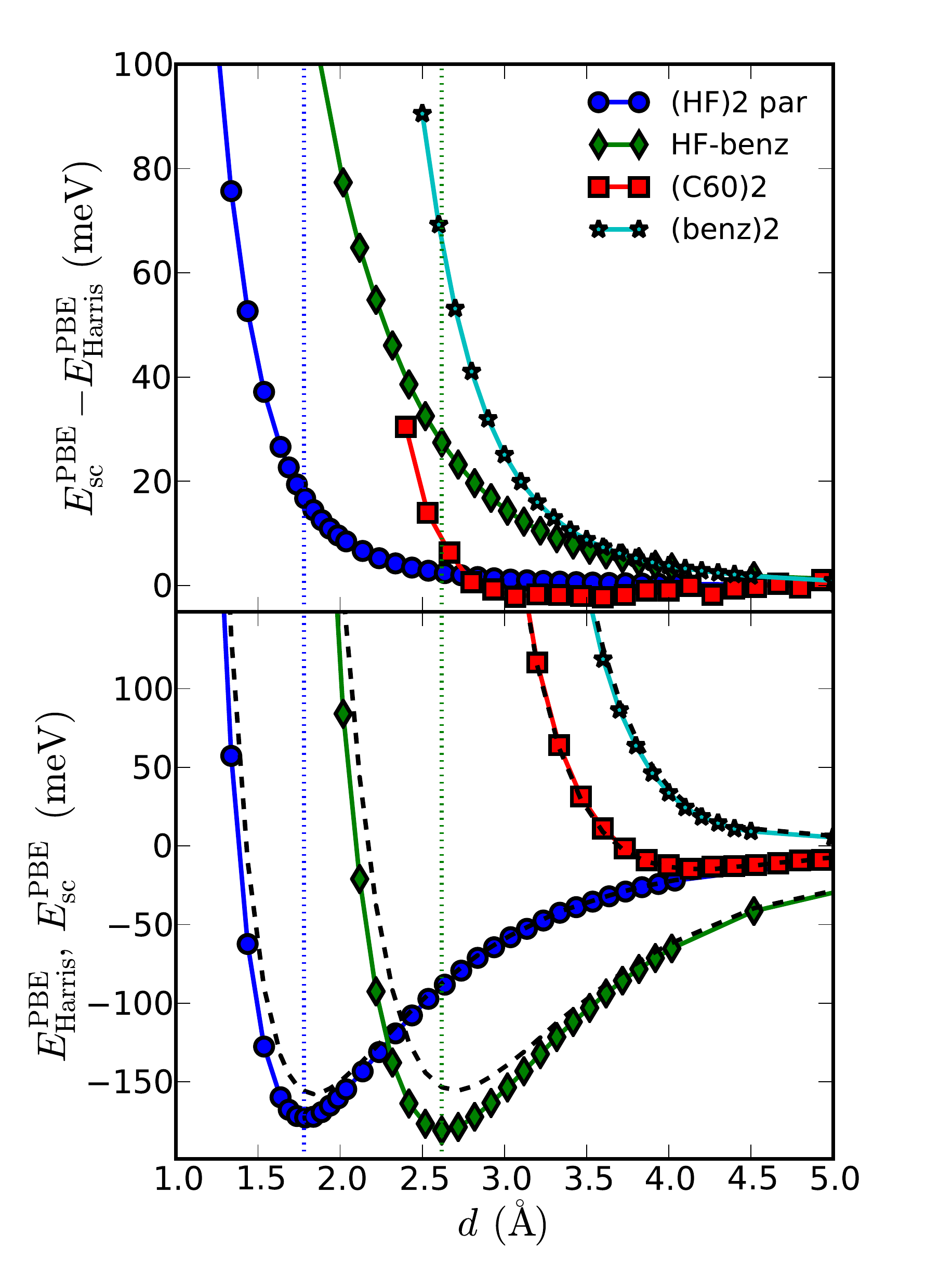}
\caption{A comparison between the regular Harris scheme 
and self-consistent (sc) Kohn-Sham 
calculations for four different molecular pair configurations identified in the upper panel,
all systems investigated in the PBE version of GGA.  These are (left to right, top
to bottom) the HF dimer in parallel configuration, 
the HF-benzene pair with the H of HF pointing towards the benzene 
center, the parallel benzene dimer, and the C60 dimer with the benzene 
rings facing each other.  
The middle panel shows the difference between the interaction energy in 
the sc and Harris PBE calculation.
The lower panel shows (full curves) the interaction curves using the 
Harris functional and (dashed curves) the sc result.
The abscissa label $d$ denotes the separation between the closest 
atoms in separate molecules.
The two curves involving the highly polar HF exhibit a non-negligible binding.
The dotted vertical lines indicate the GGA-minimum.
Among the investigated systems, the Harris estimate
gives the largest overestimate (16\%) for HF-benzene interaction curve.
}
\label{fig:GGA_dimer}
\end{figure}

Four molecular pairs, depicted in the upper panel of 
Fig.~\ref{fig:GGA_dimer}, have been chosen for our comparison between 
the sfd scheme and regular DFT calculations.
The first is a hydrogen fluoride (HF) dimer in parallel configuration. 
This configuration is not the optimal one,\cite{HFdimer,HFstructure} 
but is here chosen as a representative for systems with large dipole-dipole interactions.
The second is a molecular configuration where the hydrogen of HF points towards the center of a benzene molecule.
Thus, one molecule has zero and the other a large dipole moment in vacuum.
The third system is a benzene dimer in parallel sandwich configuration. 
The binding in this system is dominated by vdW (also called London dispersion) forces.
The interaction in this system is representative of dilute sparse matter system, like a gas. 
The fourth system, a dimer of C60 with hexagonal rings facing each other, is also one where the binding is dominated by vdW forces. 
But because of the large size of C60, this attraction is much stronger than for the benzene dimer. This system is therefore more representative of compact molecular complexes that arise in bulk sparse matter.

\subsection{Regular Harris scheme for GGA-PBE calculations}

We first describe and illustrate the Harris scheme as 
it is used for GGA calculations.
Obviously, the use of a GGA will not generally succeed in reproducing structural
properties of typical sparse, weakly interacting molecular systems. 
Nevertheless, it is instructive to illustrate that
the GGA Harris scheme 
is still generally able of faithfully reproducing the sc GGA calculations, including 
the sparse-matter GGA limitations.

Figure~\ref{fig:GGA_dimer} compares the interaction curves for the four 
different molecular pairs as obtained with
DFT and the Harris scheme using the PBE version of the GGA exchange-correlation functional. 
Only the HF dimer and the HF-benzene pair show an appreciable binding of 
respectively 173 and 180~meV using the Harris scheme for PBE and 158 and 156 meV 
using sc PBE calculations. 
For the parallel HF dimer system, which is dominated by dipole-dipole 
interactions, the Harris calculation overestimates the binding energy by 
9\% compared to regular GGA DFT calculations. 
For the HF-benzene system, where one of the molecules is highly polar and 
the other is not, the scheme overestimates the binding energy by 17\%.

The discrepancies between the two methods can be understood from the 
significant dipole moment induced by the binding. At optimal separation 
(in the selected configurations), a dipole of 0.12~e{\AA} is induced 
for the HF dimer, while one of 0.15~e{\AA} is induced for the HF-benzene pair. 
These induced dipole moments are comparable to the dipole moment of 
the HF molecule itself (0.39~e{\AA}). 
It is clear that molecular pairs involving one or more HF molecule(s) 
serve as tough tests for the feasibility of the Harris functional scheme. 

For systems dominated by the vdW forces, the discrepancies are 
difficult to assess without including the effect of non-local correlation. 
We will therefore make this assessment in the next subsection. 

\subsection{Systems dominated by vdW attraction}

\begin{figure}[t!]
\includegraphics[width=0.5\textwidth]{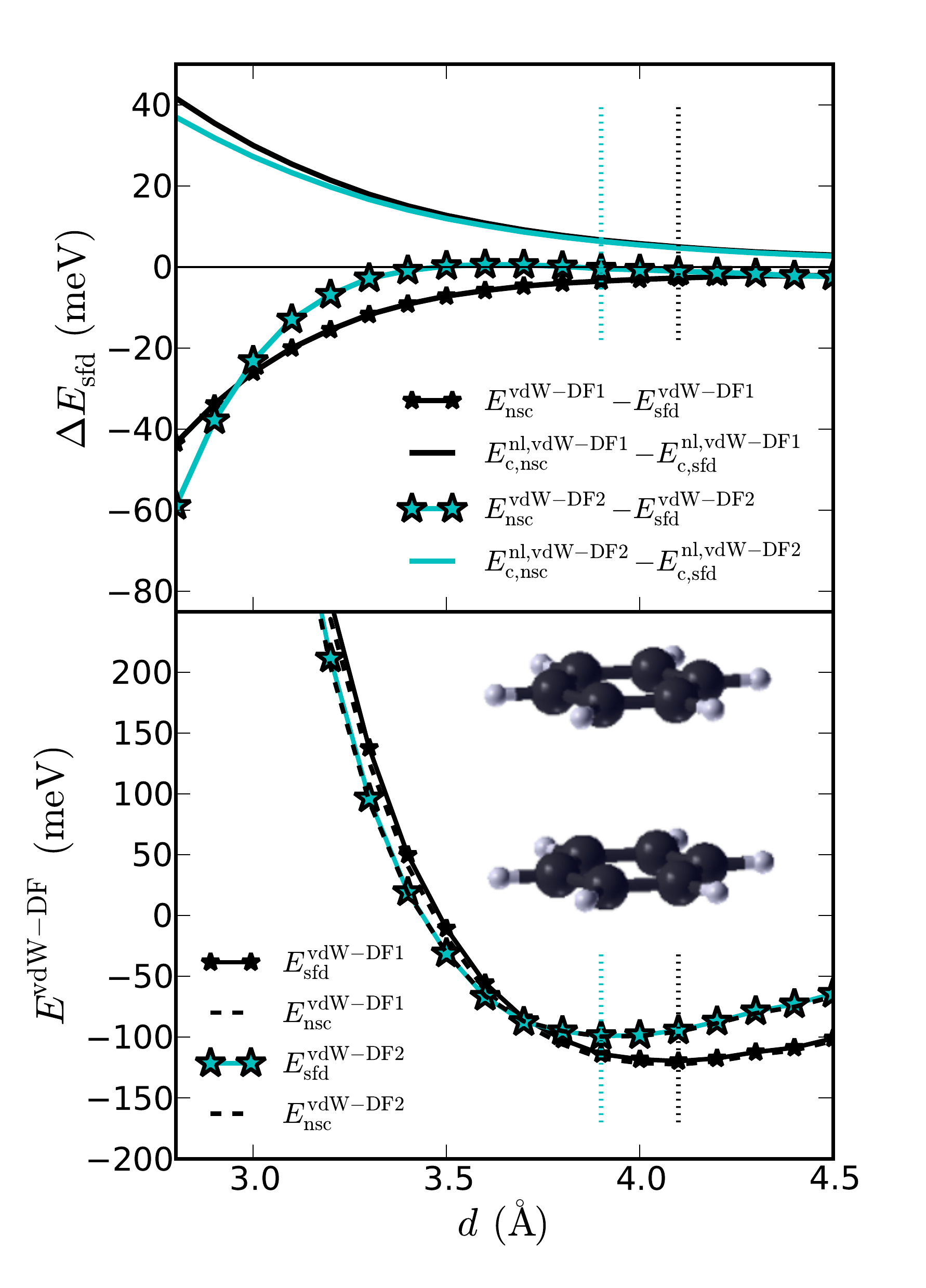}
\caption{Comparison of the sfd-vdW-DF scheme for a benzene dimer and 
results obtained in nsc-vdW-DF evaluation.
The top panel compares the
differences between these two procedures as they arise for the total energy of vdW-DF1 (black curve with star dots) and vdW-DF2  (cyan/light curves with star dots);
and the difference in the nonlocal correlation (two upper curves).
The bottom panel compares the sfd-vdW-DF1
(black star-dotted curves) and sfd-vdW-DF2 (cyan/light star-dotted 
curves) against nsc-vdW-DF results (dashed curves).
The somewhat larger shift (for most separations) in
the non-local correlation energy when using the sfd scheme compared to 
the total vdW-DF energy, in particular for vdW-DF2, indicates a partial 
cancellation of the density sensitivity of the nonlocal and semilocal components.}
\label{fig:benz_dimer}
\end{figure}

The benzene dimer is a typical organic system bound by vdW 
forces. Since this system is weakly bonded, we can expect charge transfer 
to be small and thus the vdW-DF Harris scheme, and more generally the sfd
framework, to be well suited to describe the system. 

Figure~\ref{fig:benz_dimer} shows the comparison between the sfd
and nsc vdW-DF calculations for the benzene dimer. The difference 
between the dashed and the full curves in the lower panel is barely 
distinguishable.
At binding separation the sfd result is 2\% below the nsc result for 
vdW-DF1 and merely 0.4\% for vdW-DF2. 

The upper panel of Fig.~\ref{fig:benz_dimer} shows that the non-local 
energy is somewhat affected by using the frozen density 
$n_\sfd=n_1 + n_2$ in place of the one determined 
with a full GGA calculation, $n_\nsc$.
It also reveals that there is some error cancellation 
between these shifts and the combined shifts in the other terms: 
the shifts obtained with the sfd scheme overestimate the non-local 
interaction energy, while the magnitude of the binding energy is underestimated. 
This trend is opposite to that exhibited for all four systems in the 
proper GGA Harris scheme, shown in Fig.~\ref{fig:GGA_dimer}.
For vdW-DF2 this error cancellation is close to exact
in a fairly wide region around the binding separation. 
For shorter separation between the molecules, corresponding to a larger 
density overlap, the absolute difference between schemes increases, as 
does the magnitude of the repulsive wall between molecules. 

\begin{figure}[t]
\includegraphics[width=0.5\textwidth]{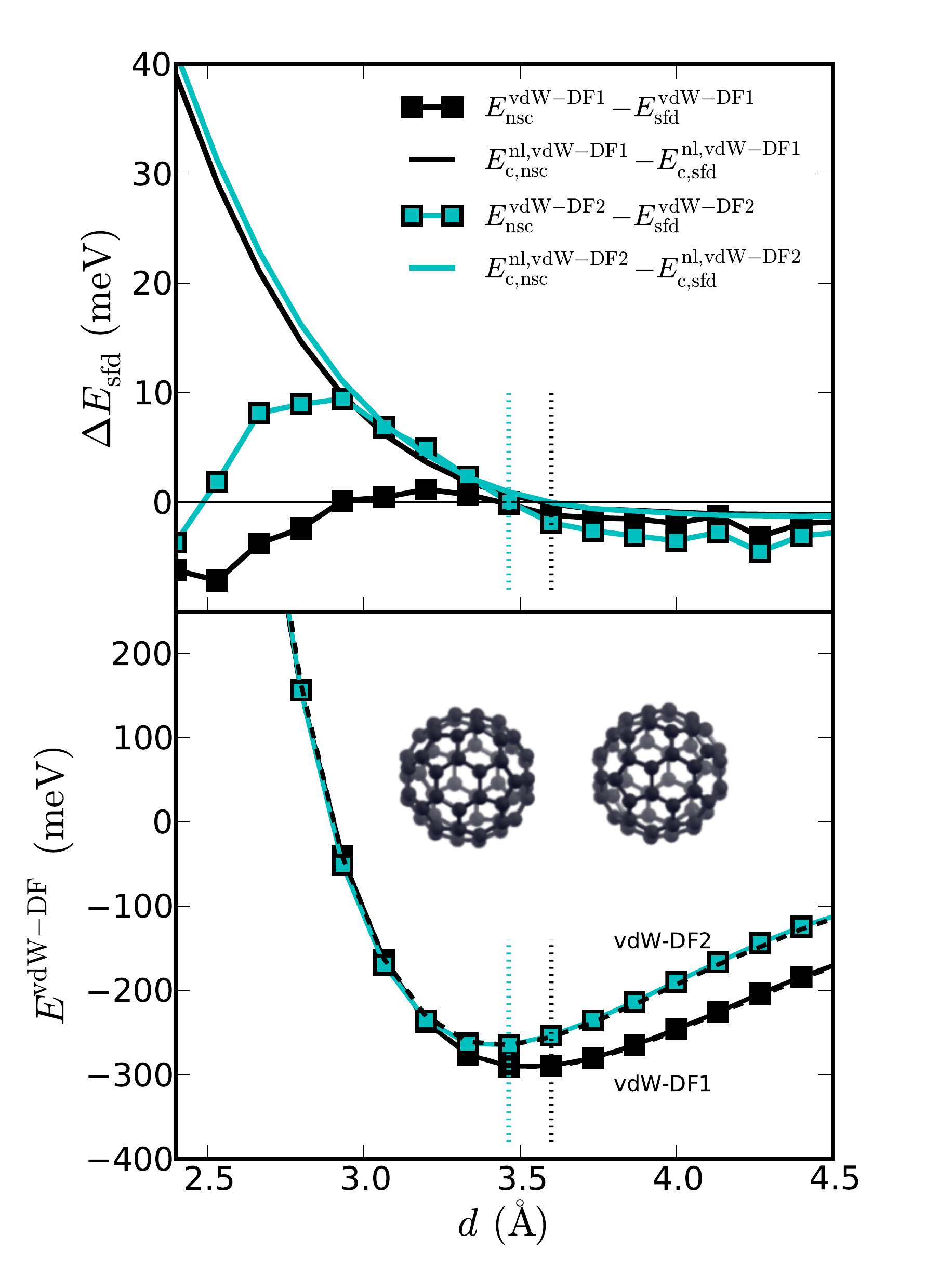}
\caption{The sfd-vdW-DF description of binding in the C60 dimer, 
legends and details as described in Fig.~\ref{fig:benz_dimer}.}
\label{fig:C60_dimer}
\end{figure}

Figure~\ref{fig:C60_dimer} compares the two methods for a C60 dimer in 
the same fashion as for the benzene dimer. 
In this case the sfd-vdW-DF1 underestimates the binding energy by as 
little as 0.2\%, while sfd-vdW-DF2 is spot on (within about 0.05 meV). 
This striking coincidence (arising from error cancellation) is likely 
fortuitous since the results are 
similar, but not this similar, in other regions of the interaction curve. 
For the C60 interaction curve, the sfd-vdW-DF calculations in some regions 
overestimate and in other regions underestimate the interaction energy. 

The benzene and C60 dimer calculations indicate that the sfd scheme is 
an appropriate method to accelerate the evaluation of interaction energies 
in systems dominated by vdW interactions. 

\subsection{System with large induced charge: HF interacting with benzene}

\begin{figure}[t]
\includegraphics[width=0.5\textwidth]{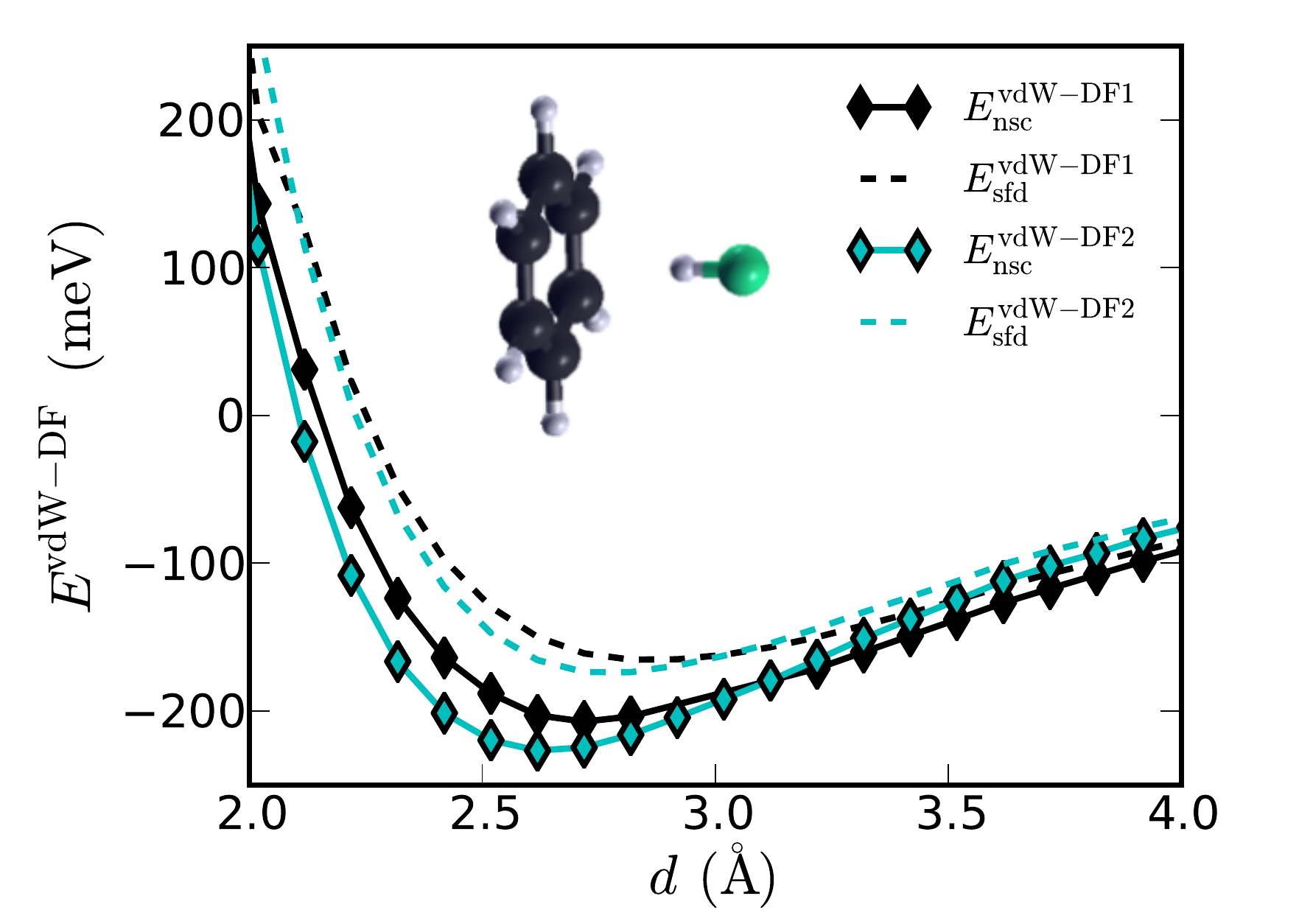}
\caption{The sfd-vdW-DF description of binding between a HF molecule 
and benzene. Legends and details as in the lower panel of Fig.~\ref{fig:benz_dimer}.
The sizable discrepancy between the sfd and sc calculation in the 
PBE calculations, shown in Fig.~\ref{fig:GGA_dimer}, carries through 
to the case of (sfd and nsc) vdW-DF calculations. The non-local correlation has only 
a tiny effect on the discrepancy.}
\label{fig:HF_benz}
\end{figure}

In the HF-benzene system the vdW forces contribute to the binding 
alongside electrostatic effects. The nsc-vdW-DF2 
predicts a binding energy of 174~meV compared to 
that of 155~meV with the sc PBE calculations (in Fig.~\ref{fig:GGA_dimer}).

Figure~\ref{fig:HF_benz} shows the results of the interaction curves 
obtained with sfd and nsc vdW-DF calculations. 
For this system, we also find that the vdW-DF2 calculation produces a 
larger binding energy than the vdW-DF1, which is opposite to the case 
for the benzene and for the C60 dimer. 
This switching of order is related to the fact that vdW-DF2 in general 
has a less repulsive exchange account~\cite{rPW86} and a less attractive 
non-local correlation account.\cite{H2benchmark} 
Since the smaller size of this system decreases the magnitude of the 
non-local correlation, it shifts the balance between the repulsive and attractive terms. 

The difference between the sfd and nsc vdW-DF calculations increases to as much 
as 30\% for vdW-DF2, compared to 17\% for the sfd and sc PBE calculations. The 
discrepancy is somewhat smaller for vdW-DF1. 
Note that this discrepancy arises mostly from the shift in the 
$\Delta E_{xc}^{\vdw,0}$ term and not from the non-local correlation, 
which contributes with 4~meV in the opposite direction of the total shift of $-60$ meV. 

Our results indicate that the increased inaccuracy of the sfd scheme for 
polar systems arises primarily from short-ranged effects.  Thus, the 
inaccuracy may be reduced by starting from 
densities generated with revPBE exchange for sfd-vdW-DF1 and 
in the same vein PW86r for sfd-vdW-DF2. 

\subsection{The nsc-vdW-DF approximation}

We note that the leading-order difference (\ref{eq:densitynscvdWDFexpress})
between sc- and nsc-vdW-DF total-energy results is 
nominally \textit{linear\/} in the density change,
$\delta n_{\scc} = n^{\vdW}_{\scc}(\mathbf{r})-n^g_{\scc}(\mathbf{r})$. 
The nsc-vdW-DF calculations---while often very successful---need 
therefore not always be as robust as a regular Harris vdW-DF scheme would be. 

On the other hand, the regular nsc-vdW-DF 
approach\cite{langrethjpcm2009} does have a mechanism for including some 
of the electron density rearrangement that arises from Pauli exclusion 
or from the formation of more traditional types of bonding (those that a 
GGA does capture). 

Fig.~\ref{fig:HF_benz} shows that keeping such charge adjustments can be important for 
systems where at least one fragment has a large static polarization.  
Generally, we expect the nsc-vdW-DF approach to be more accurate except 
in cases with very weak intermolecular interactions. Of course, the only 
way to resolve the difference would be to perform a fully sc
vdW-DF calculation.\cite{Thonhauser,Soler,Gulans} The sc-vdW-DF is now 
becoming standard procedure for medium to large systems (system sizes 
approaching a thousand atoms). However, a performance testing comparing 
$E_\nsc^\vdW$ and $E_\sfd^\vdW$ against $E_\scc^\vdW=E_\KS^\vdW$
is for technical reasons beyond the scope of this paper (as discussed 
in Section II).

\subsection{An organic-molecular testing  of sfd-vdW-DF performance}

Figs.~\ref{fig:S22_dimer} and \ref{fig:milky} and Table \ref{tab:S22}
presented a summary of the further  
assessment we have performed of the accuracy of the sfd-vdW-DF scheme
for the S22 benchmark suite.\cite{S22benchmark} 
Here we provide some additional details. 

\begin{table*}
  \begin{ruledtabular}
\caption{Interaction energies for pairs of small molecules from the S22 dataset.
Quantum chemistry (QC) results from Ref.\ \protect\onlinecite{S22QC},
the selfconsistent (sc) vdW-DF1 and vdW-DF2 results are from Ref.\ \onlinecite{vdWDF2}. 
All energies in meV/dimer. \label{tab:S22}}
\begin{tabular}{rlrrrrrrr}
\# & Duplex & $E_\sfd^\DFone$& $E^\DFone_\nsc$& $E^\DFone_\scc$
& $E_\sfd^\DFtwo$& $E^\DFtwo_\nsc$& $E^\DFtwo_\scc$& QC \\
  1 & Ammonia dimer & 118 & 111 & 115 & 136 & 126 & 134 & 137 \\
  2 & Water dimer & 186 & 171 & 185 & 220 & 201 & 218 & 218 \\
  3 & Formic acid dimer & 736 & 680 & 690 & 817 & 745 & 766 & 815 \\
  4 & Formamide dimer & 610 & 560 & 587 & 684 & 619 & 655 & 699 \\
  5 & Uracil dimer &  807 & 749 & 767 & 876 & 805 & 832 & 897 \\
  6 & 2-pyridoxine -- 2-aminopyridine &  671 & 623 & 639 & 728 & 661 & 687 & 737 \\
  7 & Adenine -- thymine & 652 & 619 & 609 & 716 & 661 & 660 & 726 \\
  8 & Methane dimer & 38 &  39 &  36 &  29 &  30 &  30 &  23 \\
  9 & Ethene dimer &  70 &  67 &  64 &  65 &  61 &  65 &  65 \\
  10 & Benzene -- methane &  72 &  70 &  68 &  64 &  62 &  63 &  63 \\
  11 & Benzene dimer (slip-parallel) &  136 & 141 & 136 & 120 & 124 & 123 & 114 \\
  12 & Pyrazine dimer & 189 & 188 & 185 & 178 & 177 & 177 & 182 \\
  13 & Uracil dimer (stacked) & 414 & 396 & 403 & 412 & 391 & 402 & 422 \\
  14 & Indole -- benzene (stacked) & 231 & 232 & 206 & 234 & 199 & 197 & 199 \\
  15 & Adenine -- thymine (stacked) &  466 & 456 & 461 & 467 & 457 & 466 & 506 \\
  16 & Ethene -- ethine &  74 &  68 &  69 &  74 &  67 &  70 &  65 \\
  17 & Benzene -- water &  142 & 125 & 124 & 148 & 126 & 129 & 143 \\
  18 & Benzene -- ammonia &  104 & 97 &  94 &  101 & 91 &  92 &  101 \\
  19 & Benzene -- HCN &  194 & 162 & 166 & 198 & 159 & 170 & 197 \\
  20 & Benzene dimer (T-shape) &  120 & 115 & 113 & 108 & 101 & 105 & 118 \\
  21 & Indole -- benzene (T-shape) & 240 & 210 & 214 & 230 & 192 & 206 & 243 \\
  22 & Phenol dimer & 267 & 252 & 254 & 290 & 270 & 279 & 307
\end{tabular}
\end{ruledtabular}
\end{table*}

Table \ref{tab:S22} presents the calculated numbers of our comparison of 
sc-, nsc-, and sfd-vdW-DF1 and -vdW-DF2 results for interaction energies in 
the S22 set of molecular dimers. 
These interaction energies are all evaluated at the binding distance
(identified in Ref.~\onlinecite{vdWDF2}) that minimizes respectively the 
sc-vdW-DF1 and sc-vdW-DF2 interaction-energy variation. 
The quantum chemistry computations are from Ref.~\onlinecite{S22QC}.
For an actual S22-benchmarking of various
sparse-matter DFT methods one should provide a full binding
energy curve for each of the computational approaches. 
Here, our purpose is merely to complement our analysis based on binding 
curves of illustrative special cases with statistics 
for the S22 set of dimers that
are seen as typical of organic-molecular interaction problems.

We note that system 1--7 can be labeled hydrogen-bonding dominated,
while 8--15 can be labeled dispersion dominated, and the remainder mixed.
By studying the table, it becomes clear that sfd tends to compare well
with nsc results for dispersion-dominated systems, while the biggest
discrepancies arise among the systems dominated by hydrogen bonds.
This observation agrees well with our analysis based on
Figs.\ \ref{fig:GGA_dimer}--\ref{fig:HF_benz}.

Figure \ref{fig:milky} conveys an overview and feeling for the quality 
of the sfd calculations compared to nsc and sc calculations. 
Together, the figure and table show that nsc and sc calculations are 
very similar. This is reassuring considering the fact that they are also 
based on different codes. 
As earlier discussed the sfd results compare well to the nsc results. 
Further, the inaccuracy introduced is overall smaller than the difference 
between sc-vdW-DF1 and QC results, while the inaccuracy is about equal to 
the difference between QC and vdW-DF2 results. vdW-DF2 has better 
performance for the S22 data set than vdW-DF1. 
The inaccuracy introduced by using sfd does not necessarily make results 
compare worse to QC.

\section{Discussion}

\subsection{On vdW-bonding effects on electron dynamics}

We begin with an interesting aside, noting the implications of our 
formal analysis on the expected error in nsc-vdW-DF and in sfd-vdW-DF,
Eqs.\ (\ref{eq:densitynscvdWDFexpress}) and (\ref{eq:densityHnscvdWDFexpress}). 

It is known that the inclusion of nonlocal correlation and vdW 
forces often gives rise to indirect bandstructure 
effects\cite{Graphane,BNbandstructure} because the vdW binding changes the 
morphology and hence the local environment for the electron 
dynamics.  However, it is also interesting to identify  
conditions where one can also expect direct vdW bandstructure effects, that is, 
electron-dynamics changes that arise when---for given structure---the 
effective potential is changed from a GGA to the vdW-DF functional form.  

The error
estimate (\ref{eq:densitynscvdWDFexpress}) allows us to identify conditions for
expecting the vdW-bonding to affect the bandstructure and, more generally, the
electron dynamics.\cite{Graphane,BNbandstructure}
A clear difference in sc and nsc vdW-DF total energies is
required in order for such direct vdW-DF bandstructure 
effects to emerge.  

Meanwhile, there now exists a significant experience with using vdW-DF 
for sparse matter systems and the calculations have shown that there is, 
\textit{in practice}, often only limited differences in nsc-vdW-DF 
and sc-vdW-DF calculations.\cite{Thonhauser,langrethjpcm2009} It follows
that one must, in general, expect that direct bandstructure effects 
typically are small.

\subsection{Overall assessment of the sfd-vdW-DF calculation scheme} 

The speed up gained when using the sfd calculations with 
\textsc{Dacapo} are substantial yet somewhat modest. Computational costs 
are reduced by 40\% and 55\% for the benzene and C60 dimer respectively, 
when using standard cutoffs with a minimal number of bands. Considering 
that this software usually requires about 20 electronic iterations 
to converge at these system sizes (but more for large systems), this gain 
is less than one might anticipate.\cite{harris} 

However, we should consider that standard software like \textsc{Dacapo} 
(that we here use) has been subjected to intense efforts to optimize its 
ability to simultaneously solve the problem of charge relaxation 
and determination of the KS eigenvalues. What formally constitutes
Harris calculations in that code are today primarily used to obtain 
accurate values for the KS eigenvalues.\cite{HarrisCoding}  
We do not desire such enhanced accuracy for an actual sfd-vdW-DF study.

The fact that the here-proposed sfd-vdW-DF is still faster than
nsc-vdW-DF (Sec.~VI) is therefore promising. Furthermore, since the 
performance, documented here, is excellent for many molecular systems, 
there is room for more compromise on accuracy.
We believe that the present results motivate the approach to be further
evaluated in forthcoming studies.\cite{HarrisCoding}

\subsection{Towards as fast biomolecular mapping of vdW interactions in biomolecular systems}

We are ultimately interested in vdW-DF computational studies for large-scale interaction problems where 
there are relevant speed ups to be gained by \textit{not\/} seeking the fully sc 
density (as described either in a full GGA  or in a full vdW-DF study). The acceleration must 
come from minimizing the cost of DFT calculations of the steric-hindrance or kinetic-energy 
repulsion effects (as this is the large-system bottleneck).

We note that a parallel study, Ref.~\onlinecite{dnaMapping}, pursues a closely
related computational strategy and \textit{begins} a first-principle
DFT mapping of the morphology variation on the vdW attraction in a DNA dimer
by first focusing on an efficient evaluation of $E_c^\nl[n_{\sfd}^g]$. The DNA 
dimer system is there taken as an example of a typical large-scale biomolecular 
interaction problem. 

The vdW-DF exploration 
that is proposed in Ref.~\onlinecite{dnaMapping} 
bypasses the need for performing the expensive computations of the kinetic-energy 
repulsion term (which, in any case, is not relevant outside the binding regime that extends to
a nearest-atoms separation of about 4.0 {\AA}, Ref.~\onlinecite{MolCrys}). 
The first-principle vdW-DF survey of DNA attraction thus achieves
a dramatic speed up but can still (for relevant large-molecular system interaction geometries)
be supplemented by adding the other components of the Harris-type
sfd-vdW-DF scheme (\ref{HtypevdWDFDefined}).

\section{Summary and outlook}
To accelerate large-scale vdW-DF characterization of biomolecular systems it seems useful 
to adapt the ideas of the Harris scheme\cite{harris,foulkes,nikulin,gordonkim} 
as is indicated and explored here.
A Harris-type approximation which works reasonably well for describing the kinetic-energy 
effects of forming covalent bonds of atoms in some molecules should have a good chance of describing 
the simpler kinetic-energy repulsion (steric hindrance) of molecules in supramolecular systems. 
 
Here, we have put this expectation to the test. 
Our results indicate that this scheme is promising for describing 
supramolecular systems bonded primarily by vdW forces. 
However, if one or more fragments are highly polar, this comes at the 
cost of accuracy.
 
This paper is also supplemented by a related publication\cite{dnaMapping} 
which presents a vdW-DF study that maps out the nonlocal correlation of 
large biopolymers within the presented sfd-vdW-DF scheme.

The pair of papers suggest a possible computational strategy for the study of
binding in large supramolecular systems. 
The suggestion is to \textit{begin\/} the structure 
and interaction-morphology search by essentially cost-free evaluations of the $E_c^\nl$ variation.
That $E_c^\nl$ step is available simply from relatively cheap calculations of fragment electron densities. 
One can in turn search for relevant binding motifs, given the 
linking to the here-proposed and 
tested sfd-vdW-DF scheme.
This strategy eventually leads to a complete vdW-DF
estimate of the variation in total interaction energy. 
The strategy exists as an alternative to implementing a real-space vdW-DF version in a code that realizes
genuine order-$N$ scaling for large systems.

Having established the promise of the sfd scheme for systems bound 
by vdW forces, the next step would be to investigate if 
the scheme can be further accelerated, in particular for large 
supramolecular systems. 
To this end, implementations for a full sfd-vdW-DF scheme for other 
DFT codes (beyond {\Dacapo}) are being tested.

\acknowledgments
The authors thank Pieremanuele Canepa 
and Timo Thonhauser for discussions and Kuyho Lee for supplying
atomic coordinates for a full range of 14 various separations for each of the
dimers in the S22 benchmark set.  Partial support from 
the Swedish Research Council (VR) (two grants) and the Chalmers Area of Advance 
`Materials' is gratefully acknowledged.  The computations were performed on 
high-performance computing resources provided by the Swedish National Infrastructure 
for Computing (SNIC) at the C3SE 
and HPC2N 
metanodes.



\begin{thebibliography}{99}
\bibitem{footnote}
The approach is sometimes also called the Harris-Foulkes scheme.
\bibitem{footnote2}
While beyond the present scope, we note that the Harris scheme with 
refinements can, for instance, be used for an approximate system 
characterization along a reaction pathway in organic systems.\protect\cite{bell}
\bibitem{HarrisCoding}
A single electronic iteration is here taken to mean an evaluation stage
during which the external potential is kept fixed.  In most DFT codes,
there are several wavefunction evaluation steps in each such electronic
iteration. When pursuing a standard sc DFT determination, the requested
accuracy on the wavefunction solution remains moderate.
For bandstructure calculations, the DFT code is formally
performing a Harris-type evaluation, but a code like
\textsc{Dacapo} then 
forces many wavefunction-evaluation steps to deliver a higher 
accuracy on the wave functions.
However, that is not the spirit of the original Harris scheme, nor
do we intend this here with the proposed sfd-vdW-DF scheme.
\bibitem{footnote3}
One can of course evaluate the KS and the Harris schemes as well as the sfd-vdW-DF approximation
for a general specification of the fragment orientation,
motif-alignment, and arbitrary choice of displacement vector.
\end{thebibliography}
\end{document}